# ExplainitAI: When do we trust artificial intelligence? The influence of content and explainability in a cross-cultural comparison

Sora Kang*
Human-Computer
Interaction+Design Lab
Seoul National University
Seoul, Republic of Korea
sorakang@snu.ac.kr

Andreea-Elena Potinteu
Perception and Action Lab
Leibniz-Institut für Wissensmedien
Tübingen, Germany
a.potinteu@iwm-tuebingen.de

Nadia Said
Applied Cognitive Psychology,
Department of Psychology
University of Tübingen
Tübingen, Germany
nadia.said@uni-tuebingen.de

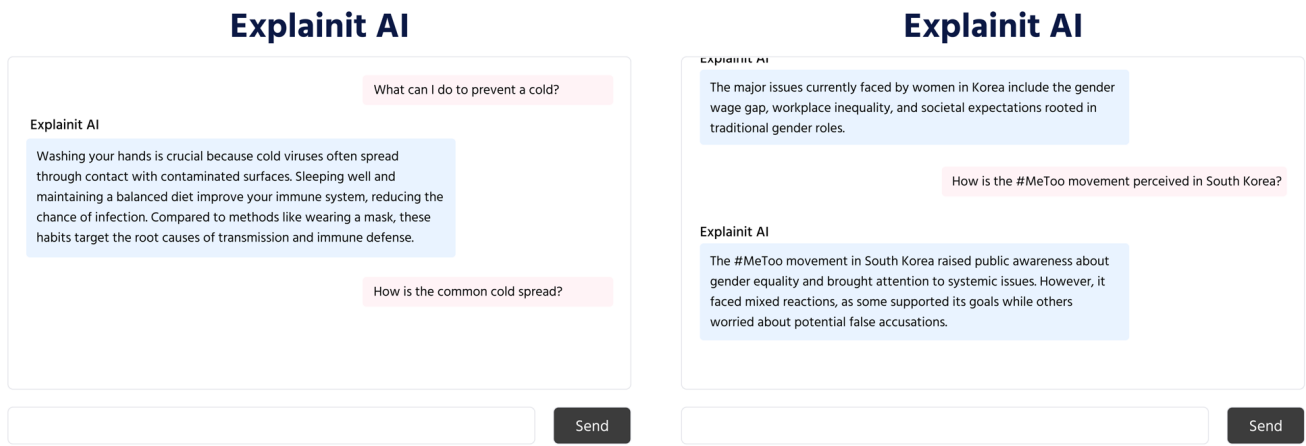

Figure 1: Explainit AI interface demonstrating high explainability (left) and low explainability (right) modes.

## Abstract

This study investigates cross-cultural differences in the perception of AI-driven chatbots between Germany and South Korea, focusing on topic dependency and explainability. Using a custom AI chat interface, ExplainitAI, we systematically examined these factors with quota-based samples from both countries ($N = 297$). Our findings revealed significant cultural distinctions: Korean participants exhibited higher trust, more positive user experience ratings, and more favorable perception of AI compared to German participants. Additionally, topic dependency was a key factor, with participants reporting lower trust in AI when addressing societally debated topics (e.g., migration) versus health or entertainment topics. These perceptions were further influenced by interactions among cultural context, content domains, and explainability conditions. The result highlights the importance of integrating cultural and contextual nuances into the design of AI systems, offering actionable insights for the development of culturally adaptive and explainable AI tailored to diverse user needs and expectations across domains.

# 1 Introduction

Artificial Intelligence (AI) technologies are transforming multiple facets of human life, from healthcare and entertainment to policymaking and governance [12, 17, 31]. These systems promise efficiency and innovation but also raise questions about trust, reliability, and cultural acceptance. As the adoption of AI becomes increasingly global, understanding how individuals from different cultural contexts perceive these technologies is critical to their successful integration and ethical deployment.

South Korea and Germany provide compelling case studies for exploring cultural differences in AI perception. Korea, a global leader in technology and AI adoption, exhibits a societal tendency toward optimism about technological advancements [14, 15, 25]. By contrast, Germany, with its strong emphasis on privacy, data security, and ethical considerations, often demonstrates a more cautious approach to emerging technologies [1, 35]. These contrasting attitudes offer a unique opportunity to investigate how cultural contexts shape trust in AI systems.

Another crucial factor in AI perception is the domain of application [2, 33]. Health applications, for instance, are often seen as beneficial, but this perception is context-dependent and varies by country and audience. In Korea, positive perceptions of health-related AI were prevalent, but this finding may not generalize to other cultural contexts [25]. Similarly, while laypeople tended to view health-related AI positively, experts expressed more skepticism [29]. In Germany, skepticism is particularly pronounced for AI applications in the media domain, largely due to concerns over data security [32]. While existing research has examined the context of AI applications, the specific impact of content on perceptions of AI has received little attention [24]. Addressing this gap, our study investigates how the nature of AI application content influences user perceptions.

Lastly, explainability, or the degree to which an AI system can provide transparent and comprehensible reasoning for its outputs, has emerged as a pivotal factor in user trust [34]. Prior research has shown that high levels of explainability enhance trust and acceptance, especially in high-stakes contexts such as healthcare [6, 8]. Despite this, the interplay between explainability, cultural context, and application domain remains underexplored.

This study aims to bridge these gaps by examining cross-cultural differences in AI perceptions, the influence of topic-specific content, and the role of explainability. By investigating these dimensions using a conversational AI, ExplainitAI, we seek to provide actionable insights for designing culturally adaptive and trustworthy AI systems.

# 2 Related Work

This section reviews prior research on cultural differences in AI trust, the role of content sensitivity and the impact of explainability on user trust, providing a foundation for understanding how these factors influence AI perception.

## 2.1 Cross-Cultural Perceptions of AI

There is an abundance of research showing that culture-differences play an important role regarding trust in AI [3, 16, 36], risk-opportunity perception [5, 14], and acceptance of AI applications. Generally, AI is perceived as more beneficial in East Asia than in Western countries [5]. A study comparing eight countries found that concerns about AI were notably lower in Korea and India, for instance [14]. That finding was further supported by Kim et al. (2022), showing that risk perception of AI was lower for Korean tweets compared to English tweets [15]. Similarly, a cross-cultural overview highlights reasons why technological acceptance (robots, algorithms, and AI) is generally higher in East Asia than in the West [37]. Based on this, we expect higher ratings for trust, positive user experiences, and opportunity perception among Korean participants compared to German participants.

## 2.2 Content Sensitivity in AI Perception

Previous research highlights the significance of AI application context in shaping attitudes, risk-opportunity perception, and usage [28, 32, 33]. A survey showed that public support for AI was higher for domains like medicine compared to art [33]. Furthermore, a study found that risk-opportunity ratings differed depending on the application contexts (transport, medical, psychological, and media) [32]. Another study demonstrated that participants' willingness to use AI varied across five different application domains [28].

This study explores topic dependency by varying the level of content contentiousness—how debated a topic is in society. This allows us to examine the impact of AI-provided content on users' trust, experience ratings, and risk-opportunity perception, independent of context. Research on this is limited, though one study found that task-specific content (e.g., therapy vs. diagnosis in healthcare) influenced perceptions of AI's capability and benevolence [24].

To our knowledge, we are the first to vary the level of content contentiousness by comparing three different domains: entertainment, health, and politics. For entertainment, we selected well-known topics in both countries (e.g., Squid Game Series, iPhone 15 Pro Max, League of Legends). For the health domain, we selected well-known, non-controversial, and less emotionally charged topics, avoiding highly debated issues. Examples include alcohol consumption, cancer check-ups, and common colds. Lastly, for the political topics, we chose topics, which are controversial in both countries: L.G.B.T.Q., Migrants, Women's Rights/#MeToo, and unification.

While one could assume that receiving information about a highly debated topic from an artificial agent might be viewed as impartial and thus, more trustworthy, we argue that peoples' tendency to anthropomorphize artificial agents [9, 10] might lead to the opposite perception: Artificial agents, chatbots, might be viewed as human-like and thus perceived as biased regarding certain topics. We expect that participants who received information about political topics will view the AI more critically than those who received information about the health topics. We expect that participants who received information about entertainment have the least critical view of AI.

## 2.3 AI Explainability and User Trust

The aim of explainable AI is to enhance peoples' understanding of the functionality of AI [4, 7, 22], by including step-by-step explanations on how a certain decision was made, or enhancement of transparency such as allowing users to further interact with the AI asking follow-up questions [11, 26]. Furthermore, contextual information, such as providing background details, clarifications, or the rationale for recommendations, has been shown to improve users' trust and comprehension by aligning AI outputs with their expectations and knowledge frameworks [19, 21].

Given that explainability is linked to higher trust in AI [34], we expect participants who receive more contextual information to have a more positive perception of AI compared to those who receive less.

## 3 Explainit AI

ExplainitAI is a conversational chatbot developed for this study using GPT-4o model, designed to explore user perceptions of AI across cultural contexts, topics, and different levels of explainability while restricting responses to relevant topics to align with the study's objectives (see Figure 1). The system's architecture integrates advanced natural language processing capabilities with a data retrieval mechanism to enable accurate, context-sensitive responses.

### 3.1 System Design and Functionality

*3.1.1 Explainability Levels.* ExplainitAI operates under two explainability conditions-High and Low Explainability. These levels are achieved through instruction tuning, ensuring consistent and contextually appropriate responses for each condition.

- **High Explainability**: Responses include detailed reasoning, step-by-step logic, and contextual information. The chatbot explains its reasoning process, highlights key factors, and provides examples when relevant.
- **Low Explainability**: Responses are concise, providing direct answers with minimal context or justification.

*3.1.2 Prompt Design and Instruction Tuning.* ExplainitAI uses carefully designed prompts to align responses with study objectives. System prompts specify the explainability level (high or low), guiding the chatbot to provide either detailed reasoning or concise outputs. Embedded examples ensure consistent responses across interactions. A query filtering mechanism ensures relevance by limiting topics to politics, health, and entertainment, while off-topic queries receive polite refusals with suggestions for reframing. It is designed to operate in each Korean and German, as the experiment is conducted in these respective languages.

*3.1.3 Data Integration and Vector Store.* To ensure accurate and relevant responses, ExplainitAI incorporates a custom vector store with curated datasets in politics, health, and entertainment, collected by both Korean and German scholars. This structure enables efficient information retrieval and enhances the contextual precision of the chatbot's outputs.

### 3.2 Chat Interface Development

The user interface for ExplainitAI was developed to support real-time interaction, allowing users to engage in natural conversations with immediate, tailored responses. Its intuitive design, incorporating familiar messaging aesthetics, ensures accessibility and ease of use for diverse demographics. To support research objectives, the interface includes a real-time logging system that records interactions while maintaining user anonymity, ensuring both data integrity and privacy. By integrating GPT-4o model with curated datasets and explainability adjustments, the interface provides context-sensitive, topic-relevant responses. This design makes ExplainitAI an effective tool for investigating how cultural context, topics, and explainability influence user perceptions of AI.

## 4 Study Design

The study was pre-registered at *https://aspredicted.org/6yqn-8scf.pdf* and approved by the ethics board of the Leibniz-Institut für Wissensmedien, Tübingen. The study was designed as a 2 (country) x 2 (explainability: high vs low) x 3 (topic: politics, health, entertainment) x 4 (subtopic) between-subject design.

### 4.1 Participants

Participants were recruited via the panel provider Bilendi & respondi. We aimed to collect a total of $N$ = 300 participants (quota-based sample regarding age and gender with $n$ = 150 from Germany and $n$ = 150 from Korea) to ensure statistical power. A total of $N$ = 304 participants were collected. For ethical reasons, participants had the opportunity to withdraw their data at the end of the study, thus the remaining sample size was $N$ = 297, with $n$ = 153 participants from Germany and $n$ = 144 from Korea. For Germany, the mean age was $M$ = 49.59 years ($SD$ = 14.91 years), with $n$ = 75 female participants and $n$ = 78 male participants. Regarding education most participants either reported having a university degree (30.06%) or a secondary school certificate (28.76%). For Korea, the mean age was $M$ = 46.55 years ($SD$ = 15.19 years), with $n$ = 74 female participants and $n$ = 70 male participants. Regarding education most participants (56.25%) reported having a university degree.

### 4.2 Procedure

Participants provided informed consent and completed a demographic questionnaire before interacting with ExplainitAI. They were then presented with topic-specific prompts (five questions) related to one subtopic from the three domains (politics, health, or entertainment). Their task was to prompt ExplainitAI with those questions. Depending on whether participants were in the high or low explainability condition, they either received a very short answer (low) or a longer answer with additional information (high) on the topic. To ensure that participants did not skip the task, the "next" button was disabled for 40 seconds. Furthermore, to assess engagement, they were asked a multiple-choice question about the topic afterward. Post-interaction surveys assessed trust, perceived utility, and risk-opportunity perceptions. At the end of the study, participants were fully debriefed about the study's aim and received information about the sources for each topic. In a final step, participants provided consent for data processing and had the option to withdraw their data. The entire process was conducted in the participant's preferred language, either Korean or German.

### 4.3 Measures

*Topics.* Participants received 5 topic-specific questions, which they should prompt the AI with. In total there were 12 topics, four within each category: politics (L.G.B.T.Q., Migrants, Women's Rights/#MeToo, unification), entertainment (Squid Game Series,

iPhone 15 Pro Max, League of legends, Lotte World Adventure/Europa Park), health (alcohol consumption, cancer check-up, common cold, tick bites). Trust scores, AI user experience ratings, and risk-opportunity perception are averaged for participants within the subtopics. That is, there were $n = 52$ participants from Korea and $n = 46$ participants from Germany in the politics group, $n = 45$ participants from Korea and $n = 53$ participants from Germany in the entertainment group, and $n = 47$ participants from Korea and $n = 54$ participants from Germany in the health group.

*Control Questions.* To control whether participants engaged with the content presented by ExplanitAI, they had to answer one multiple choice question after interacting with the AI. Overall, about 73% of participants answered the question correctly.

*Trust in AI.* Participants' trust in AI was assessed with 8 items from the Trust Scale for Explainable AI (TXAI) [13]. Participants had to rate statements about AI (e.g., "I trust the AI. I think the AI works well.") on a 5-point scale from ("1 = completely agree", to "5 = completely disagree"). Note that for better comparison with the other measures, ratings were reverse coded for the analysis. Cronbach's alpha for the trust ratings was $\alpha = .84$. The trust score was calculated as the average of the responses to the items.

*AI User Experience Ratings (UER).* User experience ratings were assessed with 8 adjectives (emotional, social, analytical, disturbing, private, pleasant, useful, innovative) on a 7-point Likert scale [28]. Participants were asked "In your opinion, to what extent do the following adjectives apply to artificial intelligence (AI)?" and had to rate "1 = applies not at all" to "7 = completely applies". Cronbach's alpha for the user experience ratings was $\alpha = .78$. The UER score was calculated as the average of the responses to the items.

*Risk-Opportunity Perception.* Participants' risk-opportunity perception was measured with 6 items (health, elderly care, entertainment, art, society, political topics) on a 7-point scale ranging from "risk" to "opportunity" ("Do you see the use of artificial intelligence (AI) as a source of information in the following areas more of a risk or an opportunity?"). Cronbach's alpha for the risk-opportunity items was $\alpha = .76$. The risk-opportunity perception score was calculated as the average of the responses to the items.

## 5 Findings

As pre-registered we conducted MANOVAs for all three independent variables (country, topic, explainability). For the hypothesis and the corresponding pre-registered analysis please see: *https://aspredicted.org/6yqn-8scf.pdf*. Additionally, we conducted exploratory linear regressions, controlling for age, gender, education, political attitudes, and correct responses to the control question. Furthermore, we conducted Box's M tests for equal covariance matrices, Bartlett tests of homogeneity of variances, and Shapiro-Wilk normality tests to check whether the assumptions for running MANOVAs and ANOVAs are met. While the assumption of equal covariance matrices was violated for country as independent variable ($X^2 (6) = 35.89$, $p < .001$), for the other two independent variables topics and explainability, the assumption was met ($X^2 (12) = 17.12$, $p = 0.145$; $X^2 (6) = 7.08$, $p = 0.314$ [38]). For ANOVAs, Shapiro-Wilk tests were significant, though ANOVAs are generally robust to normality violations. Bartlett tests indicated homogeneity violations for trust and user experience labels but not for risk-opportunity scores (for country). Consequently, we also conducted Kruskal-Wallis tests (see 5.1).

### 5.1 Cross-Cultural Perceptions of AI

We conducted a MANOVA comparing trust, user experience ratings, and risk-opportunity perception between Germany and Korea. Results showed a significant main effect of the variable country: $F(1, 295) = 11.77$, $p < .001$; with a medium effect size $\eta^2$ *(partial)* $= 0.11$, 95% CI [0.05, 1.00]. Conducting single ANOVAS revealed that for all three dependent variables ratings were significantly higher for Korea compared to Germany.

- AI trust: $F(1, 295) = 28.25$, $p < .001$; $\eta^2 = 0.09$, 95% CI [0.04, 1.00].
- AI user experience: $F(1, 295) = 30.11$, $p < .001$; $\eta^2 = 0.09$, 95% CI [0.05, 1.00].
- Risk-opportunity: $F(1, 295) = 10.54$, $p = .001$; $\eta^2 = 0.03$, 95% CI [8.44e-03, 1.00].

Furthermore, we conducted Kruskal-Wallis tests, which were significant in all three cases: $X^2 (1) = 25.11$, $p < .0001$ (trust), $X^2 (1) = 34.77$, $p < .0001$ (user experience), $X^2 (1) = 8.30$ $p = .004$ (risk-opportunity). When additionally conducting linear regressions, results remained the same regarding the differences in ratings for Germany and Korea.

- AI trust: $F(6, 290) = 7.515$, $p < .0001$, $R^2 = .135$, $R^2$ adjusted $= .12$, $b = 0.37$, $t(290) = 4.78$, $p < .0001$.
- AI user experience: $F(6, 290) = 8.061$, $p < .0001$, $R^2 = .143$, $R^2$ adjusted $= .125$, $b = 0.58$, $t(290) = 5.33$, $p < .0001$.
- Risk-opportunity: $F(6, 290) = 3.502$, $p < .01$, $R^2 = .068$, $R^2$ adjusted $=.048$, $b =0.405$, $t(290) = 3.170$, $p < .01$.

### 5.2 Content Sensitivity in AI Perception

We conducted a MANOVA comparing trust, user experience ratings, and risk-opportunity perception between the different topics (health, entertainment, politics). There was no significant main effect: $F(2, 294) = 1.04$, $p = 0.398$; $\eta^2$ *(partial)* $= 0.01$, 95% CI [0.00, 1.00]. When additionally conducting linear regressions, results showed that participants had significantly lower trust ratings after prompting the AI about political topics compared to the entertainment topics: $F(7, 289) =3.656$, $p < .001$, $R^2 = 0.081$, $R^2$ adjusted $=0.059$, $b = -0.199$, $t(289) = -2.148$, $p = .0325$ (see Figure 2, left panel). Lastly, when adding country as interaction variable, while the MANOVA did not yield a significant interaction effect, $F(2, 291) = 1.45$, $p = 0.194$; $\eta^2$ *(partial)* $= 0.01$, 95% CI [0.00, 1.00]; linear regressions showed a significant interaction between topics and country for risk-opportunity perception ratings: $F(10, 286) = 2.681$, $p = .0037$, $R^2 = .086$, $R^2$ *adjusted* $=.054$, $b = 0.68$, $t(286) = 2.28$, $p = .0235$. More specifically, while participants from Germany and Korea did not differ in their risk-opportunity ratings after receiving information about political topics, participants from Germany perceived AI less as an opportunity for the health topic compared to participants from Korea (see Figure 2, right panel).

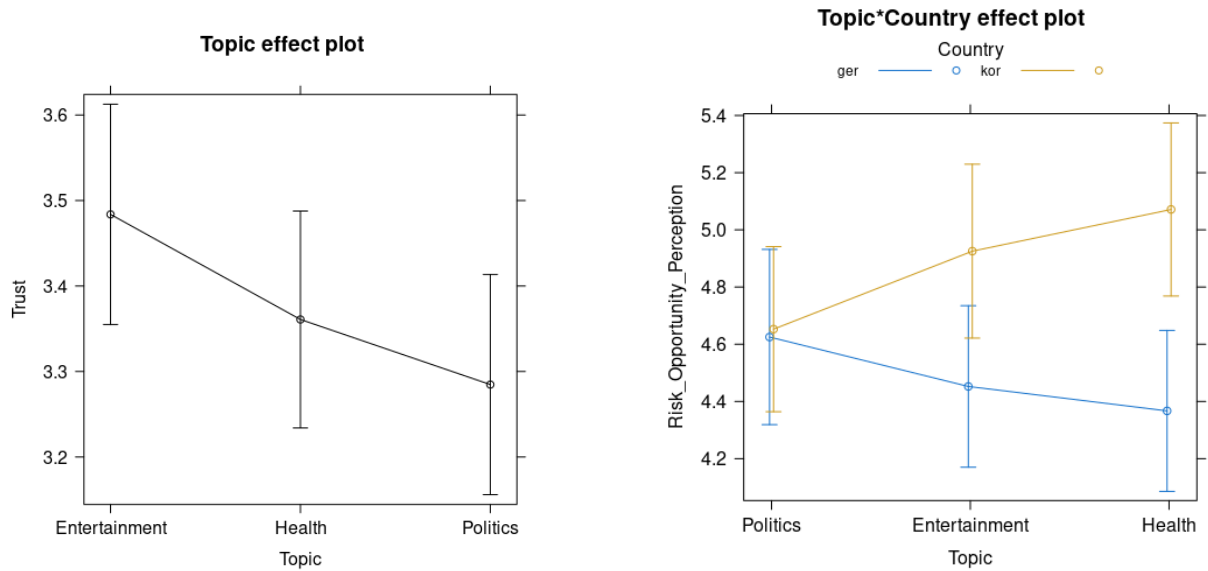

Figure 2: The figure displays the effects extracted from the linear regressions for AI trust ratings by topic (left) and topic-country interactions (right). Error bars represent 95% confidence intervals.

### 5.3 AI Explainability and User Trust

We conducted a MANOVA comparing trust, user experience ratings, and risk-opportunity perception for the two explainability conditions (high vs low). There was no significant main effect: F(1, 295) = 2.29, p = 0.079; $\eta^2$ (partial) = 0.02, 95% CI [0.00, 1.00]. There was no significant main effect: $F(1, 295) = 2.29$, $p = 0.079$; $\eta^2$ (partial) = 0.02, 95% CI [0.00, 1.00]. When additionally conducting linear regressions for the three dependent variables, there was no significant effect as well. However, results were hinting towards higher trust for the high explainability condition with: $F(6, 290) = 3.99$, $p < .001$, $R^2 = .076$, $R^2$ adjusted =.057, b = 0.132, $t(290) = 1.755$, $p = .0803$. Lastly, when adding topics as interaction variable – while the MANOVA yielded no significant effects – linear regressions showed a significant interaction effect for the user experience ratings: $F(10, 286) = 2.58$, $p = .005$, $R^2 = .08267$, $R^2$ adjusted =.05059, b = -0.64, $t(286) = -2.393$, $p = .0174$. More specifically, participants' user experience ratings were lower in the health category for the high explainability condition compared to the political and entertainment categories (see Figure 3).

## 6 Discussion and Future Work

We investigated the impact of cross-cultural differences in AI perceptions, the influence of topic-specific content, and the role of explainability on trust in AI, AI user experience ratings, and risk-opportunity perception. We thus aimed at further understanding what factors influence trust in AI, AI user experience ratings, and risk-opportunity perception, which are key prerequisites for technology adoption [18, 20, 23, 28, 30, 32, 34].

### 6.1 Cross-Cultural Perceptions of AI

Regarding cross-cultural differences, we did find that for all three measures Korean participants had significantly higher ratings (for both types of analysis, MANOVA and linear regressions). More specifically, German participants reported lower trust, higher risk perception, and lower positive user experience compared to participants from Korea. This is in line with previous literature investigating differences in AI perception for countries like Japan, Korea, the UK, and India [14, 16, 36], showing that generally acceptance of new technologies is higher in East Asian countries. Our findings thus further emphasize the importance of taking cultural differences into account when introducing new technologies. Given that our results also show the importance of explainable AI (see below), in countries where trust in AI is lower, efforts to increase transparency, explainability, and user control might be crucial for acceptance. Conversely, in cultures with higher trust in AI, companies may focus on enhancing user experience and integration.

### 6.2 Content Sensitivity in AI Perception

When investigating the impact of topic-specific content on AI perception, participants reported significantly lower trust in AI when prompted with political topics compared to entertainment topics, even after controlling for age, gender, education, political attitudes, and correctly answered control questions. Furthermore, an interaction between content type and country emerged for risk-opportunity perception: while Korean participants generally viewed AI as more of an opportunity than German participants, this difference was particularly pronounced for health topics compared to political topics.

While the AI was framed as a general chatbot, the variability in trust and risk-opportunity perception, shows that content is a key factor in AI perception dynamics. A possible explanation could be that people perceive chatbot-type AIs more as a conversation partner and less as an advanced search engine, thus transferring human-like traits (e.g., opinions) to the AI. However, research on the context of AI implementation also shows variability in AI perception [28, 31, 32], where medical AIs are perceived differently to security ones for example. Thus, it could be that for certain topics, people simply refuse to rely on information if it is opposed to one's own beliefs, independent of who provides this information [27]. This implies that even when an AI provides reliable information, people would be biased in their perception, by their own beliefs on the content rather than the AI. To untangle the underlying mechanisms that lead to such differences in perception based on the content provided by the AI, future studies should include measures like attitudes regarding the respective topics, perceived anthropomorphism, and knowledge about AI.

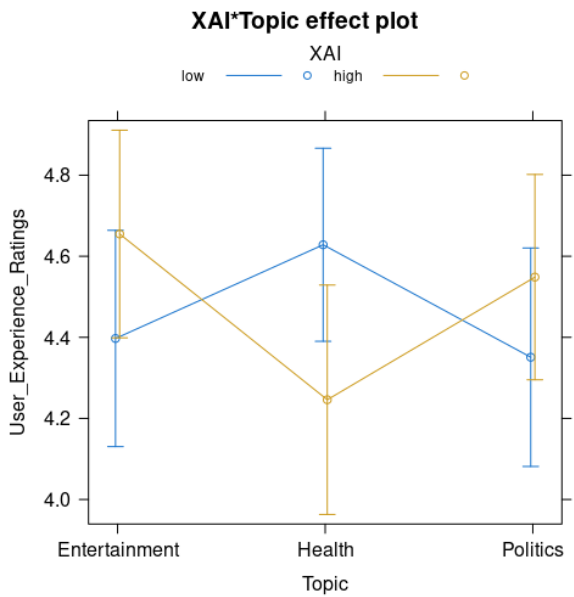

**Figure 3: The figure displays the effects extracted from the linear regressions for the interaction between explainability (high vs. low) and topics. Error bars represent 95% confidence intervals.**

## 6.3 AI Explainability and User Trust

Lastly, we investigated the influence of the amount of contextual information on AI perception. While additional information did not significantly impact user experience ratings or risk-opportunity perception, results suggested a potential effect on trust in AI. Specifically, participants reported descriptively higher trust ratings in the high explainability condition ($M = 3.45$, $SD = 0.62$) compared to the low explainability condition ($M = 3.31$, $SD = 0.71$). This effect may be small, and our sample size might not have been sufficient to detect it, warranting further exploration in future studies. While user experience ratings were higher in the high explainability condition for the entertainment and political topics, the health topic showed the opposite pattern, with higher ratings in the low explainability condition. One possible explanation is that for health topics, users may prefer to receive detailed explanations from a physician rather than an AI, which could influence their perceptions of the AI in this context. Taken together, our results corroborate previous results in the area of explainable AI that showed that among other factors, additional information can enhance users' trust in AI applications [19, 21].

## 7 Limitations And Conclusion

While one strength of our study is the use of quota-based samples from two countries, ensuring generalizability, the sample sizes may not have been sufficient to detect small effects. Regarding explainability, we only varied one factor—the amount of contextual information provided. Future research should consider varying other factors, for example providing participants with the opportunity to ask follow-up and clarification questions, thus enhancing user-AI engagement. The research shows that perceptions of AI are based on an intricate interplay between cultural differences, content provided by the AI, and design of the AI itself (e.g., explainability). By introducing content as a novel aspect to AI research, our findings provide actionable insights for the development of culturally adaptive and explainable AI systems.

## References

[1] Hartmut Aden. 2021. Privacy and Security: German Perspectives, European Trends and Ethical Implications. In Ethical Issues in Covert, Security and Surveillance Research. Emerald Publishing Limited, 119–129. https://doi.org/10.1108/S2398-601820210000008009

[2] Saleh Afroogh, Ali Akbari, Evan Malone, Mohammadali Kargar, and Hananeh Alambeigi. 2024. Trust in AI: Progress, Challenges, and Future Directions. https://doi.org/10.48550/arXiv.2403.14680

[3] Vishakha Agrawal, Serhiy Kandul, Markus Kneer, and Markus Christen. 2023. From OECD to India: Exploring cross-cultural differences in perceived trust, responsibility and reliance of AI and human experts. https://doi.org/10.48550/arXiv.2307.15452

[4] Plamen P. Angelov, Eduardo A. Soares, Richard Jiang, Nicholas I. Arnold, and Peter M. Atkinson. 2021. Explainable artificial intelligence: an analytical review. WIREs Data Min & Knowl 11, 5 (September 2021), e1424. https://doi.org/10.1002/widm.1424

[5] Aaron J. Barnes, Yuanyuan Zhang, and Ana Valenzuela. 2024. AI and culture: Culturally dependent responses to AI systems. Curr Opin Psychol 58, (August 2024), 101838. https://doi.org/10.1016/j.copsyc.2024.101838

[6] Michael Da Silva. 2023. Explainability, Public Reason, and Medical Artificial Intelligence. Ethic Theory Moral Prac 26, 5 (November 2023), 743–762. https://doi.org/10.1007/s10677-023-10390-4

[7] Arun Das and Paul Rad. 2020. Opportunities and Challenges in Explainable Artificial Intelligence (XAI): A Survey. https://doi.org/10.48550/arXiv.2006.11371

[8] Christopher Ifeanyi Eke and Liyana Shuib. 2024. The role of explainability and transparency in fostering trust in AI healthcare systems: a systematic literature review, open issues and potential solutions. Neural Comput & Applic (December 2024). https://doi.org/10.1007/s00521-024-10868-x

[9] Nicholas Epley, Adam Waytz, and John T. Cacioppo. 2007. On seeing human: a three-factor theory of anthropomorphism. Psychol Rev 114, 4 (October 2007), 864–886. https://doi.org/10.1037/0033-295X.114.4.864

[10] Karin van Es and Dennis Nguyen. ?Your Friendly Ai Assistant?: The Anthropomorphic Self-Representations of Chatgpt and its Implications for Imagining Ai. AI and Society, 1–13. https://doi.org/10.1007/s00146-024-02108-6

[11] David Gunning, Mark Stefik, Jaesik Choi, Timothy Miller, Simone Stumpf, and Guang-Zhong Yang. 2019. XAI—Explainable artificial intelligence. Sci. Robot. 4, 37 (December 2019), eaay7120. https://doi.org/10.1126/scirobotics.aay7120

[12] Pavel Hamet and Johanne Tremblay. 2017. Artificial intelligence in medicine. Metabolism 69, (April 2017), S36–S40. https://doi.org/10.1016/j.metabol.2017.01.011

[13] Robert R. Hoffman, Shane T. Mueller, Gary Klein, and Jordan Litman. 2023. Measures for explainable AI: Explanation goodness, user satisfaction, mental models, curiosity, trust, and human-AI performance. Front. Comput. Sci. 5, (February 2023). https://doi.org/10.3389/fcomp.2023.1096257

[14] Patrick Gage Kelley, Yongwei Yang, Courtney Heldreth, Christopher Moessner, Aaron Sedley, Andreas Kramm, David T. Newman, and Allison Woodruff. 2021. Exciting, Useful, Worrying, Futuristic: Public Perception of Artificial Intelligence in 8 Countries. In Proceedings of the 2021 AAAI/ACM Conference on AI, Ethics, and Society, July 21, 2021. ACM, Virtual Event USA, 627–637. https://doi.org/10.1145/3461702.3462605

[15] Jang Hyun Kim, Hae Sun Jung, Min Hyung Park, Seon Hong Lee, Haein Lee, Yonghwan Kim, and Dongyan Nan. 2022. Exploring Cultural Differences of Public Perception of Artificial Intelligence via Big Data Approach. In HCI International 2022 Posters, Constantine Stephanidis, Margherita Antona and Stavroula Ntoa (eds.). Springer International Publishing, Cham, 427–432. https://doi.org/10.1007/978-3-031-06417-3_57

[16] Jarosław Kozak and Stanisław Fel. 2024. How sociodemographic factors relate to trust in artificial intelligence among students in Poland and the United Kingdom. Sci Rep 14, 1 (November 2024), 28776. https://doi.org/10.1038/s41598-024-80305-5

[17] Ralf T. Kreutzer and Marie Sirrenberg. 2020. Fields of Application of Artificial Intelligence—Financial Services and Creative Sector. In Understanding Artificial Intelligence: Fundamentals, Use Cases and Methods for a Corporate AI Journey, Ralf T. Kreutzer and Marie Sirrenberg (eds.). Springer International Publishing, Cham, 211–224. https://doi.org/10.1007/978-3-030-25271-7_8

[18] JeungSun Lee, Min-Kyu Kwak, and Seong-Soo Cha. 2021. The Effect of Motivated Consumer Innovativeness on Perceived Value and Intention to Use for Senior Customers at AI Food Service Store. Journal of Distribution Science 19, 9 (2021), 91–100. https://doi.org/10.15722/jds.19.9.202109.91

[19] Zachary C. Lipton. 2016. The Mythos of Model Interpretability. https://doi.org/10.48550/ARXIV.1606.03490

[20] Lars Meyer-Waarden and Julien Cloarec. 2022. "Baby, you can drive my car": Psychological antecedents that drive consumers' adoption of AI-powered autonomous vehicles. Technovation 109, (January 2022), 102348. https://doi.org/10.1016/j.technovation.2021.102348

[21] Tim Miller. 2018. Explanation in Artificial Intelligence: Insights from the Social Sciences. https://doi.org/10.48550/arXiv.1706.07269

[22] Brent Mittelstadt, Chris Russell, and Sandra Wachter. 2019. Explaining Explanations in AI. In Proceedings of the Conference on Fairness, Accountability, and Transparency, January 29, 2019. 279–288. https://doi.org/10.1145/3287560.3287574

[23] Farzaneh Nasirian, Mohsen Ahmadian, and One-Ki (Daniel) Lee. 2017. AI-Based Voice Assistant Systems: Evaluating from the Interaction and Trust Perspectives. AMCIS 2017 Proceedings (August 2017). Retrieved from https://aisel.aisnet.org/amcis2017/AdoptionIT/Presentations/27

[24] Ekaterina Novozhilova, Kate Mays, Sejin Paik, and James E. Katz. 2024. More Capable, Less Benevolent: Trust Perceptions of AI Systems across Societal Contexts. MAKE 6, 1 (February 2024), 342–366. https://doi.org/10.3390/make6010017

[25] Songhee Oh, Jae Heon Kim, Sung-Woo Choi, Hee Jeong Lee, Jungrak Hong, and Soon Hyo Kwon. 2019. Physician Confidence in Artificial Intelligence: An Online Mobile Survey. Journal of Medical Internet Research 21, 3 (March 2019), e12422. https://doi.org/10.2196/12422

[26] P Jonathon Phillips, Carina A Hahn, Peter C Fontana, Amy N Yates, Kristen Greene, David A Broniatowski, and Mark A Przybocki. 2021. Four principles of explainable artificial intelligence. National Institute of Standards and Technology (U.S.), Gaithersburg, MD. https://doi.org/10.6028/NIST.IR.8312

[27] Chanthika Pornpitakpan. 2004. The Persuasiveness of Source Credibility: A Critical Review of Five Decades' Evidence. Journal of Applied Social Psychology 34, 2 (2004), 243–281. https://doi.org/10.1111/j.1559-1816.2004.tb02547.x

[28] Andreea Elena Potinteu, Daniela Renftle, and Nadia Said. 2023. What Predicts AI Usage? Investigating the Main Drivers of AI Use Intention over Different Contexts. https://doi.org/10.31234/osf.io/jvdpe

[29] Said, N., Schumacher, L. & Huff, M. (2023, September 29). Artificial Intelligence in Medicine: The Influence of Medical Expertise and Perceived Causability on Medical AI Risk and Benefit Perception. https://osf.io/preprints/psyarxiv/m5q3p

[30] Nadia Said, Andreea E. Potinteu, Irina Brich, Jürgen Buder, Hanna Schumm, and Markus Huff. 2023. An artificial intelligence perspective: How knowledge and confidence shape risk and benefit perception. Comput. Hum. Behav. 149, C (Spring 2023). https://doi.org/10.1016/j.chb.2023.107855

[31] Birgit Schippers. 2020. Artificial Intelligence and Democratic Politics. Political Insight 11, 1 (March 2020), 32–35. https://doi.org/10.1177/2041905820911746

[32] Rebekka Schwesig, Irina Brich, Jürgen Buder, Markus Huff, and Nadia Said. 2023. Using artificial intelligence (AI)? Risk and opportunity perception of AI predict people's willingness to use AI. Journal of Risk Research 26, 10 (2023), 1053–1084.

[33] Neil Selwyn, Beatriz Gallo Cordoba, Mark Andrejevic, and Liz Campbell. 2020. AI for Social Good - Australian Attitudes Toward AI and Society Report.pdf. Monash University. https://doi.org/10.26180/13159781.V1

[34] Donghee Shin. 2021. The effects of explainability and causability on perception, trust, and acceptance: Implications for explainable AI. International Journal of Human-Computer Studies 146, (February 2021), 102551. https://doi.org/10.1016/j.ijhcs.2020.102551

[35] Harrison Stewart and Jan Jürjens. 2018. Data security and consumer trust in FinTech innovation in Germany. ICS 26, 1 (March 2018), 109–128. https://doi.org/10.1108/ICS-06-2017-0039

[36] Olga Viberg, Mutlu Cukurova, Yael Feldman-Maggor, Giora Alexandron, Shizuka Shirai, Susumu Kanemune, Barbara Wasson, Cathrine Tømte, Daniel Spikol, Marcelo Milrad, Raquel Coelho, and René F. Kizilcec. 2024. What Explains Teachers' Trust of AI in Education across Six Countries? https://doi.org/10.48550/arXiv.2312.01627

[37] Kai Chi Yam, Tiffany Tan, Joshua Conrad Jackson, Azim Shariff, and Kurt Gray. 2023. Cultural Differences in People's Reactions and Applications of Robots, Algorithms, and Artificial Intelligence. Manag. Organ. Rev. 19, 5 (October 2023), 859–875. https://doi.org/10.1017/mor.2023.21

[38] Finch, H. 2005. Comparison of the performance of nonparametric and parametric MANOVA test statistics when assumptions are violated. Methodology: European Journal of Research Methods for the Behavioral and Social Sciences, 1, 1, 27–38. https://doi.org/10.1027/1614-1881.1.1.27